\documentclass[12pt]{article}
\usepackage{epsfig}
\usepackage{a4p}

\begin{document}
\def\be{\begin{equation} }
\def\ee{\end{equation} }
\def\ba{\begin{eqnarray} }
\def\ea{\end{eqnarray} }
\def\ban{\begin{eqnarray*} }
\def\ean{\end{eqnarray*} }
\def\epem{\mbox{e}^+\mbox{e}^-}
\def\eegg{\epem\to\gamma\gamma}
\def\eeggg{\epem\to\gamma\gamma(\gamma)}
\def\ggg{\gamma\gamma(\gamma)}
\def\ct{\cos{\theta}}
\def\cte{\cos{\theta^{\ast}}}
\def\pl{p_{\rm l}}
\def\pt{p_{\rm t}}
\def\g{\gamma}
\def\B{{\mathcal{B}}}
\def\O{{\mathcal{O}}}
\def\R{{\mathcal{R}}}
\def\E{{\mathcal{E}}}
\def\xmc{\left(\frac{d\sigma}{d\Omega}\right)_{\rm MC}}
\def\xb{\left(\frac{d\sigma}{d\Omega}\right)_{\rm Born}}
\def\xl{\left(\frac{d\sigma}{d\Omega}\right)_{\Lambda_{\pm}}}
\def\xq{\left(\frac{d\sigma}{d\Omega}\right)_{\rm QED+7}}
\def\xr{\left(\frac{d\sigma}{d\Omega}\right)_{\rm QED+6}}
\def\xs{\left(\frac{d\sigma}{d\Omega}\right)_{\rm QED+8}}
\def\xe{\left(\frac{d\sigma}{d\Omega}\right)_{\rm e^{\ast}}}
\def\xsn{\frac{d\sigma}{d\Omega}}
\def\ca{$I$}
\def\cb{$I\!I$}
\def\cc{$I\!I\!I$}

\begin{titlepage}
\begin{center}{\Large   EUROPEAN LABORATORY FOR PARTICLE PHYSICS
}\end{center}\bigskip
\begin{flushright}
       CERN-PPE/97-109   \\ 11 August 1997
\end{flushright}
\bigskip\bigskip\bigskip\bigskip\bigskip
\begin{center}{\LARGE\bf  
\boldmath
Multi-photon final states in $\epem$ collisions at \\[1ex]
$\sqrt{s} =$ 130 -- 172 GeV 
\unboldmath
}\end{center}\bigskip\bigskip
\begin{center}{\LARGE The OPAL Collaboration
}\end{center}\bigskip\bigskip

\begin{abstract}
The process $\eegg(\g)$ is studied using data recorded with the OPAL detector 
at LEP. The data 
sample corresponds to a total integrated luminosity of 25.38 pb$^{-1}$ taken 
at centre-of-mass energies of 130 - 172 GeV. 
The measured cross-sections agree well with the expectation from QED.  
In a combined fit using data from all centre-of-mass energies, the 
angular distribution is used to obtain improved limits on the cut-off 
parameters: $\Lambda_+ > $ 195 GeV and $\Lambda_- > $ 210 GeV (95\% CL).
In addition, limits on non-standard $\epem\g$ couplings and contact 
interactions, as well as a 95\%\ CL mass limit for an excited electron, 
$M_{\rm e^{\ast}} > 194$ GeV for an $\epem\g$ coupling 
$\kappa = 1$, are determined. 
\end{abstract}
\bigskip\bigskip\bigskip\bigskip
\bigskip\bigskip
\begin{center}{\large
(Submitted to Zeit. Phys. C)
}\end{center}
\end{titlepage}
\begin{center}{\Large        The OPAL Collaboration
}\end{center}\bigskip
\begin{center}{
K.\thinspace Ackerstaff$^{  8}$,
G.\thinspace Alexander$^{ 23}$,
J.\thinspace Allison$^{ 16}$,
N.\thinspace Altekamp$^{  5}$,
K.J.\thinspace Anderson$^{  9}$,
S.\thinspace Anderson$^{ 12}$,
S.\thinspace Arcelli$^{  2}$,
S.\thinspace Asai$^{ 24}$,
D.\thinspace Axen$^{ 29}$,
G.\thinspace Azuelos$^{ 18,  a}$
% ...
}\end{center}\bigskip
\bigskip
$^{  1}$School of Physics and Space Research, University of Birmingham,
Birmingham B15 2TT, UK
\newline
$^{  2}$Dipartimento di Fisica dell' Universit\`a di Bologna and INFN,
I-40126 Bologna, Italy
\newline
% ...
$^{  5}$Cavendish Laboratory, Cambridge CB3 0HE, UK
\newline
% ...
$^{  8}$CERN, European Organisation for Particle Physics,
CH-1211 Geneva 23, Switzerland
\newline
$^{  9}$Enrico Fermi Institute and Department of Physics,
University of Chicago, Chicago IL 60637, USA
\newline
% ...
$^{ 12}$Indiana University, Department of Physics,
Swain Hall West 117, Bloomington IN 47405, USA
\newline
% ...
$^{ 16}$Department of Physics, Schuster Laboratory, The University,
Manchester M13 9PL, UK
\newline
% ...
$^{ 18}$Laboratoire de Physique Nucl\'eaire, Universit\'e de Montr\'eal,
Montr\'eal, Quebec H3C 3J7, Canada
\newline
% ...
$^{ 23}$Department of Physics and Astronomy, Tel Aviv University,
Tel Aviv 69978, Israel
\newline
$^{ 24}$International Centre for Elementary Particle Physics and
Department of Physics, University of Tokyo, Tokyo 113, and
Kobe University, Kobe 657, Japan
\newline
% ...
$^{ 29}$University of British Columbia, Department of Physics,
Vancouver BC V6T 1Z1, Canada
\newline
% ...
$^{  a}$ and at TRIUMF, Vancouver, Canada V6T 2A3
\newline
% ...
\newline
\bigskip

\section{Introduction}

This paper reports a study of the annihilation process $\eeggg$ using data 
recorded with the OPAL detector at LEP. At LEP energies,
this is one of the few processes having negligible contributions from the weak 
interaction. Since the QED differential cross-section is precisely predicted in
theory, deviations from the expected angular distribution are a sensitive test 
for non-standard physics processes contributing to these photonic final states.

The OPAL collaboration has previously published a study of photonic final 
states, with and without missing energy, at $\sqrt{s} =$ 130 - 140 GeV
\cite{graham}. The present analysis concentrates on final states with two or 
more detected photons, but no missing transverse momentum, to study only the QED 
process. Photonic final states with missing energy have been analysed 
separately \cite{misspt}.

Any non-QED effects described by the general framework of effective Lagrangian 
theory should increase with centre-of-mass energy. 
Existing OPAL limits on deviations from QED can be improved by using 
the data at centre-of-mass energies of 161.3 GeV and 172.1 GeV. 
A small amount of data taken at 170.3 GeV is included in the 172 GeV sample. 
The corresponding integrated luminosities of these data sets are 
9.97 and 10.13 pb$^{-1}$, respectively. Since the selection criteria have changed,
previously analysed data taken at centre-of-mass energies of \mbox{130.3 GeV} 
(2.69 pb$^{-1}$) and \mbox{136.2 GeV} (2.59 pb$^{-1}$) are reanalysed here
to allow for a coherent treatment.  The 136 GeV sample includes a small amount of 
data taken at 140.2 GeV. The error on the luminosity differs slightly
for the different energies and is approximately 0.5\% .

These measurements test QED at the highest centre-of-mass energies. 
Possible deviations are conveniently parametrised by cut-off parameters
$\Lambda_{\pm}$.
A comparison of the measured photon angular distribution with the QED 
expectation leads to limits on the QED cut-off parameters $\Lambda_{\pm}$,
contact interactions ($\epem\g\g$) and non-standard $\epem\g$-couplings
as described in section 3. 
The possible effects of an excited electron, $\rm{e}^*$, which would also
change the angular distribution, are investigated. In addition, the 
possible production of a resonance X via \mbox{$\epem\to$ X$\gamma$}, followed 
by the decay \mbox{X $\to \g\g$}, is studied in the invariant mass spectrum of 
photon pairs in three-photon final states. 

The following section contains a brief description of the OPAL detector
and the Monte Carlo simulated event samples.
Section 3 describes the QED differential cross-sections for $\eeggg$, as well
as those from several models containing extensions to QED.
In sections 4 - 6 the analysis is described in detail. The results are
presented in section 7. 

\section{The OPAL detector and Monte Carlo samples}

A detailed description of the OPAL detector can be found in \cite{det}.
OPAL uses a right-handed coordinate system in which the $z$ axis is along
the electron beam direction and the $x$ axis is horizontal. The polar angle,
$\theta$, is measured with respect to the $z$ axis and the azimuthal angle,
$\phi$, with respect to the $x$ axis.
For this analysis the most important detector component is the 
electromagnetic calorimeter (ECAL) which is divided into two parts, the barrel 
and the endcaps. The barrel covers the polar angle range of $|\ct |<0.81$ and
consists of 9440 lead-glass blocks. The endcaps cover the polar 
angle range of $0.81<|\ct |<0.98$ and consist of 1132 blocks.
In this analysis, the central tracking detector is used
primarily to reject events inconsistent with purely photonic final states.
Raw hit information from the vertex drift chamber (CV) and 
jet drift chamber (CJ) is used to reject events with tracks coming from the interaction
point. CV is divided into 36 $\phi$-sectors and its inner 12 (6) axial layers 
cover an angular range of $|\ct |<0.95 (0.97)$.
CJ is divided into 24 $\phi$-sectors and covers an angular range of 
$|\ct |<0.97$ with its inner 16 layers.
Incorporated in the surrounding magnet yoke is the hadronic 
calorimeter (HCAL) covering 97\%\ of the solid angle.
The outermost detectors are the muon chambers, shielded from the interaction 
point by at least 1.3 m of iron and covering 
the polar angle range of $|\ct |<0.985$.

Different Monte Carlo samples are used to study efficiency and background.
For the signal process $\eeggg$ the RADCOR \cite{mcgen} generator is used.
It provides a $\O(\alpha^3)$ cross section up to $|\ct | = 1$ for the
photon angle. No Monte Carlo with complete fourth order is currently available.
The only program that generates four-photon final states neglects the 
mass of the electron and therefore does not correctly include photons
in the far forward range. The Bhabha process is studied with two
different programs. BHWIDE \cite{mcbh} generates both electron and positron in
the acceptance of the detector. In contrast, TEEGG \cite{mcte} allows one of them
to have very low energy or escape along the beam-pipe, in addition, one photon
is scattered into the detector. The process $\epem\to\overline{\nu}\nu\g(\g)$
is studied with NUNUGPV \cite{mcnn}. Both $\epem\to\mu^+\mu^-$ and 
$\epem\to\tau^+\tau^-$ are simulated using KORALZ \cite{mctt} and PYTHIA
\cite{mcmh}
is used for hadronic events. All samples were processed through the
OPAL detector simulation program \cite{mcdet} and reconstructed as for real data.

\section{Cross section for the process \boldmath $\eegg$ \unboldmath}

The differential cross-section for the process $\eegg$ in the relativistic
limit of lowest order QED is given by \cite{bg}:
\be
\xb = \frac{\alpha^2}{s}\;\frac{1+\cos^2{\theta} }{1-\cos^2{\theta} }.
\label{born}
\ee
where $s$ denotes the square of the centre-of-mass energy, $\alpha$ is the
electromagnetic coupling constant and $\theta$ the polar angle of one photon. 
Since the two photons cannot be distinguished the event angle is defined such 
that $\ct$ is positive. 

In Ref. \cite{drell} possible deviations from the QED cross-section for Bhabha 
and M{\o}ller scattering are pa\-ra\-me\-tri\-zed in terms of cut-off parameters. 
These parameters correspond to a short range exponential term added 
to the Coulomb potential. This ansatz leads to a modification of the 
photon angular distribution as given in Eq. (\ref{lambda}). 
\be
\xl \hspace{6.5mm}  =  \xb \; \left[
1\pm\frac{s^2}{2\Lambda_\pm^4}\sin^2{\theta} \right]
\label{lambda} 
\ee
Alternatively, in terms of effective Lagrangian theory, a gauge invariant operator may be added to 
QED. Depending on the dimension of the operator different deviations from QED 
can be formulated \cite{eboli}. 
Contact interactions ($\g\g\epem$) or non-standard $\g\epem$ couplings 
described by dimension 6, 7 or 8 operators lead to angular distributions 
with different mass scales $\Lambda$ (see Eqs. \ref{qed6} - \ref{qed8}).  
The subscripts ({\small QED+6} etc.) follow the notation in Ref. \cite{eboli}.
\ba
\xr & = & \xb \; \left[
1 + \frac{s^2}{\alpha\Lambda^4_6}\sin^2{\theta} \right]
\label{qed6} 
\\
\xq & = & \xb + \frac{s^2}{32\pi}\frac{1}{\Lambda^6_7} 
\label{qed7} 
\\
\xs & = & \xb + \frac{s^2 m_{\rm e}^2}{32\pi}\frac{1}{\Lambda^8_8} 
\label{qed8} 
\ea
The definition of Eq. (\ref{qed6}) is identical to the 
standard definition  (Eq. \ref{lambda}) if 
$\Lambda_\pm^4 = \frac{\alpha}{2}\Lambda^4_6$.
Similarly Eq. (\ref{qed7}) is equivalent to Eq. (\ref{qed8})
if $\Lambda^8_8 = m_{\rm e}^2\;\Lambda^6_7$. 
Therefore only the 
parameters of Eq. (\ref{lambda}) and (\ref{qed7}) are determined by a
fit to obtain limits on deviations from QED. The limits on the other parameters
can easily be derived from these results.
 
The existence of an excited electron ${\rm e}^*$ with an
${\rm e}^*{\rm e}\g$ coupling
would contribute to the photon production process via $t$-channel exchange.
The resulting deviation from $\xb$ depends on the ${\rm e^*}$ mass $M_{\rm e^*}$ and
the coupling constant $\kappa$ of the $\mathrm{e^* e\g}$ vertex \cite{litke}:
\ba \label{estar}
\left(\frac{d\sigma}{d\Omega}\right)_{\rm e^*} & = &
\xb + \\
\alpha^2 \Biggm\{ & & \frac{1}{2}\left( \frac{\kappa}{M_{\rm e^*}} \right)^4 
    \left( E^2 \sin^2{\theta} + M_{\rm e^*}^2 \right)
    \left(\frac{q^4}{(q^2 - M_{\rm e^*}^2)^2} + 
    \frac{q'^4}{(q'^2 - M_{\rm e^*}^2)^2} \right)  \nonumber\\
& + & 4 \left( \frac{\kappa}{M_{\rm e^*}} \right)^4
     \frac{M_{\rm e^*}^2 \; E^4 \: \sin^2{\theta}}
     {(q^2 - M_{\rm e^*}^2)(q'^2 - M_{\rm e^*}^2)} \nonumber\\
& + & \left( \frac{\kappa}{M_{\rm e^*}} \right)^2 
 \left[\frac{q^2}{q^2 - M_{\rm e^*}^2} + \frac{q'^2}{q'^2 - M_{\rm e^*}^2} 
    +  E^2 \sin^2{\theta} 
    \left(\frac{1}{q^2 - M_{\rm e^*}^2} + \frac{1}{q'^2 - M_{\rm e^*}^2}\right)
    \right]\Biggm\} , \nonumber
\ea
with the beam energy $E = \sqrt{s} /2$, $q^2 = -2E^2(1-\cos{\theta})$ and
$q'^2 = -2E^2(1+\cos{\theta})$. In the limit $ M_{\rm e^*} \gg \sqrt{s}$, the
mass is related to the cut-off parameter by 
$ M_{\rm e^*} = \sqrt{\kappa}\;\Lambda_+$.

\section{Event angle definition and radiative corrections}
\label{angdef}

For the process $\epem\to\g_1\g_2$ the polar angle $\theta$ of the event is 
defined by the angle between either of the two photons and the beam 
direction since $|\cos{\theta_1}| = |\cos{\theta_2}|$. This is a good approximation for most 
of the events under consideration, since additional photons tend to be soft.
For many events, however, there is a third energetic photon and thus
$|\ct_1| \ne |\ct_2|$ in general. Several angle definitions are possible to 
characterize an event. The following two are considered:
\ba 
\cos{\theta_{\rm av}} &=& \frac{|\cos{\theta_1}| + |\cos{\theta_2}|}{2}, 
\label{ctav} \\
\cos{\theta^{\ast}} &=& \left|\sin{\frac{\theta_1 - \theta_2}{2}}\right|
           \; {\Bigg /} \; {\sin{\frac{\theta_1 + \theta_2}{2}}},
\label{ctstar} \ea
where $\theta_1$ and $\theta_2$ are the polar angles of the most energetic 
photons. Both $\cos{\theta_{\rm av}}$ and  $\cte$ are identical to $|\ct |$ 
for two-photon final states. 
For three-photon events in which the third photon is along the beam direction,
$\theta^{\ast}$ is equivalent to the scattering angle in the centre-of-mass 
system of the two observed photons. 

\begin{figure}[p]
   \begin{center} \mbox{
          \epsfxsize=16.0cm
           \epsffile{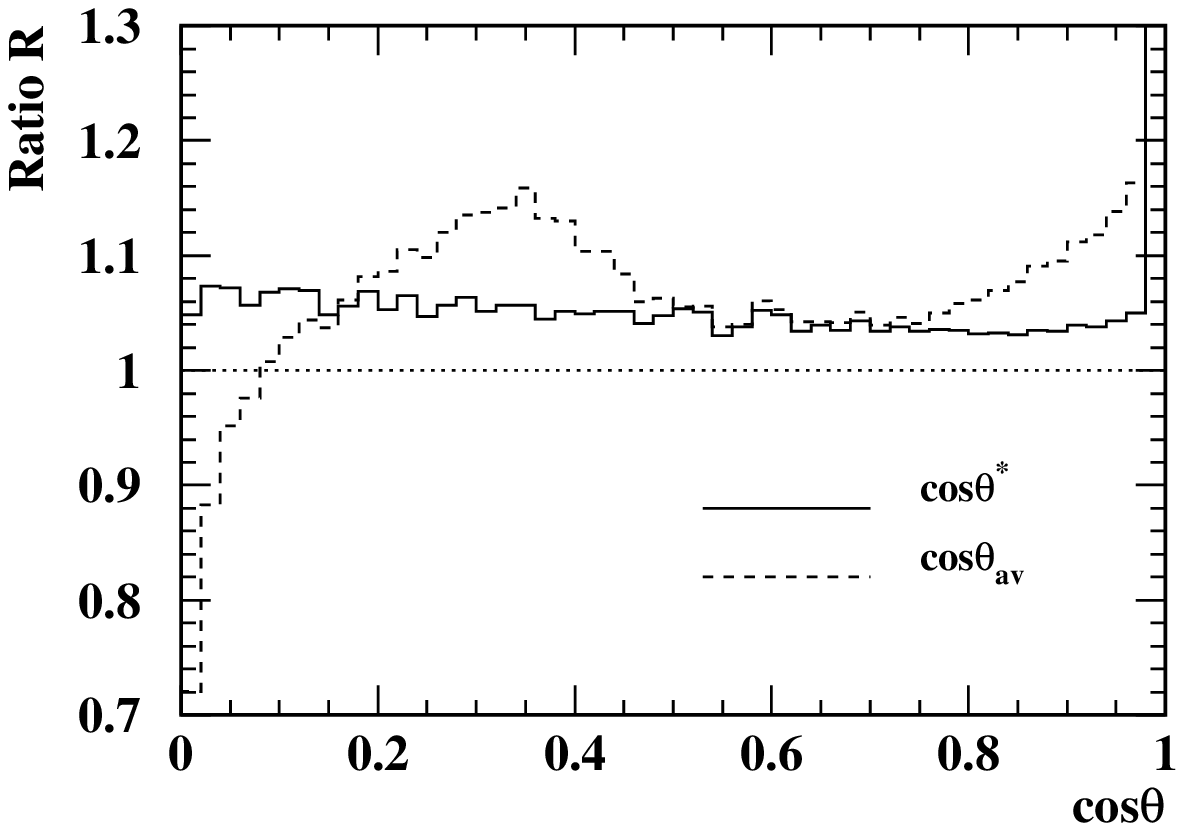}
           } \end{center}
\caption[ ]{Ratio of the 
differential cross-section for the $\eeggg$ Monte Carlo sample
relative to the Born cross-section, $\R = \xmc / \xb$. 
$\R$ is shown here for both angular definitions $\ct_{\rm av}$ (Eq. \ref{ctav}) 
and $\cte$ (Eq. \ref{ctstar}). The event angle is calculated from the two
photons with the highest generated energy.}
\label{mctest}
\end{figure}

Fig. \ref{mctest} shows the ratio of the angular distributions using
both $\cos{\theta_{\rm av}}$ and $\cte$, relative to the Born cross section as derived using 
an $\O(\alpha^3)$ $\eeggg$ Monte Carlo generator \cite{mcgen}. The event angles 
are calculated from the two photons with the highest generated energy. 
The comparison is made at the generator level, i.e. without detector simulation 
and efficiency effects. It can be seen that the distribution of 
$\cos{\theta_{\rm av}}$ (Eq. \ref{ctav}) shows large deviations from the lowest 
order (Born) distribution for much of the $\ct$ range. For this analysis 
$\cte$  (Eq. (\ref{ctstar})) is chosen because it better matches the shape of
the Born distribution over the range $\cte < 0.9$ considered in this
analysis. 
 
\section{Event selection}

Events are selected by requiring two or more clusters in the 
electromagnetic calorimeter (ECAL). A cluster is selected
as a photon candidate if it is within the polar angle range 
$|\ct| < 0.97$. The cluster must consist of at least two lead-glass blocks, 
with a combined ECAL energy deposit exceeding 1 GeV uncorrected for possible
energy loss in the material before the ECAL.
Events with a photon candidate having five or more
reconstructed clusters within a cone with a half-angle of 11.5$^{\circ}$
are rejected. This isolation criterion helps to reduce some instrumental background.

There are two major classes of background remaining to the $\ggg$ signature. 
The first can be identified by the presence of primary charged tracks. Bhabha 
events, for example, have
similar electromagnetic cluster characteristics as $\ggg$ events, 
but are normally easily distinguished by
the presence of charged tracks. The second class consists of events without 
primary charged tracks. Certain cosmic ray events and the Standard Model 
process $\epem\to\bar{\nu}\nu\g\g$ contribute to this background.

\subsection{Neutral events}
Events having only photons in the final state are classified as `neutral events'.
They should not have any charged track consistent with coming from the 
interaction point.
The rejection of all events having tracks in the central tracking chambers
CV or CJ would lead to an efficiency loss because of converted photons. 
Nevertheless, contributions from any channel with primary charged tracks 
should be reduced to a negligible level. 

To reject events with primary charged tracks while retaining efficiency
for converted photons, only the inner part of the drift chambers are considered.
First, the correlation between the observed clusters and charged hit activity 
in both drift chambers is used. Hits are counted in the 
$\phi$-sectors of CV and CJ which are geometrically associated to each 
cluster. A correlation is assigned to a cluster if there are more than a 
given number of wires with hits in the associated $\phi$-sector.
\begin{itemize}
\item A CV correlation is assigned if there are at least $m$ 
     wires with hits in the $n$ CV layers nearest to the beam-pipe 
     (denoted by $m/n$), depending on $\ct$ of the cluster: 
     \begin{center}
     \begin{tabular}[t]{r|rcl}
     Cut on $m/n$ & \multicolumn{3}{c}{$\ct$ region} \\ \hline
     6/12        &  0. &$< |\ct| <$&0.75 \\
     5/8         & 0.75&$< |\ct| <$&0.95 \\
     4/6 or 5/8  & 0.95&$< |\ct| <$&0.97   \\ 
     \end{tabular}
     \end{center}
\item A CJ correlation is assigned if there are at least 12 wires with hits
      in the inner 16 CJ layers, independent of the cluster polar angle.
\end{itemize}
Two vetoes are defined using combinations of these hit activity 
correlations in CV and CJ. A third veto tests for reconstructed charged tracks 
not correlated with either of the clusters. Any of the three vetoes rejects the 
event.
\begin{itemize}
\item The {\bf single veto} requires that both the CV and CJ 
      correlation are assigned for any cluster. 
\item The {\bf double veto} requires that for each of the highest energy clusters either 
      the CV or CJ correlation is assigned.
\item The {\bf unassociated track veto} requires that there be no reconstructed  
      track with a transverse momentum of more than 1 GeV and at least 20 hits 
      in CJ, separated by more than $10^{\circ}$ in $\phi$ from all 
      photon candidates. 
\end{itemize}

\subsection{Cosmic ray events}
A cosmic ray particle can pass through the hadronic and electromagnetic 
calorimeters without necessarily producing a reconstructed track in the 
central tracking chambers. Events of this type are rejected if 
there are 3 or more hits in the muon chambers. In the case of 1 or 2 muon hits 
the event is rejected if the highest energy HCAL cluster with at least 1 GeV 
is separated from 
each of the photon candidates by more than 10$^{\circ}$ in $\phi$.
Events are rejected if the cluster extent in $\ct$ is larger than 0.4.
This cut is primarily to reject beam halo events.

\subsection{Kinematic selection}
The event sample is divided into three classes \ca , \cb\ and \cc . The classes are 
distinguished by the number of photon candidates and the acollinearity angle 
$\zeta$, defined as $\zeta = 180^{\circ} - \xi$, where $\xi$ is the  
angle between the two highest energy clusters. Different selections are applied
to each class separately, to make use of the different kinematics. Only events
with $\cte < $ 0.9 are selected to avoid systematic errors 
due to large efficiency and radiative corrections. 

\begin{figure}[htbp]
   \begin{center} \mbox{
          \epsfxsize=16.0cm
           \epsffile{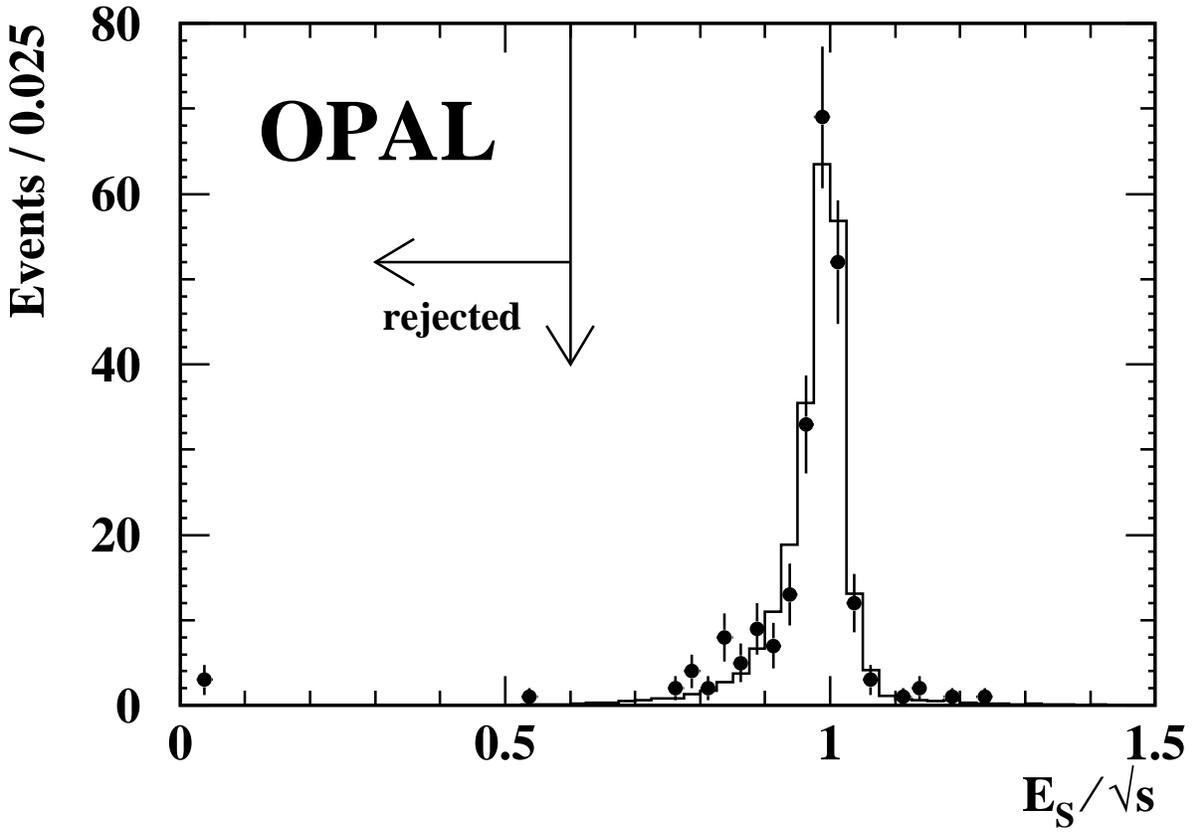}
           } \end{center}
\caption[ ]{Scaled sum of the two highest cluster energies for all events with 
an acollinearity angle $\zeta < 10^{\circ}$ (corresponding to class \ca ). 
The points with error bars 
represent the data, the histogram the $\eeggg$ Monte Carlo
expectation. The cut on this quantity is indicated.}
\label{gg}
\end{figure}

All events having an acollinearity angle $\zeta < 10^{\circ}$ (i.e. the two 
highest energy clusters are almost collinear) belong to class \ca\ independent 
of the number of photon candidates. 
For true $\eeggg$ events in this class, the sum of the two highest cluster 
energies $E_S = E_1 + E_2$ should almost be equal to
the centre-of-mass energy $\sqrt{s}$. The distribution of $E_S$ is shown 
in Fig. \ref{gg}. Events having $E_S > 0.6 \sqrt{s}$ are selected. This 
cut is 
well below the tail of the energy distribution for $\eeggg$ 
Monte Carlo events as shown in the figure. 

\begin{figure}[htbp]
   \begin{center} \mbox{
          \epsfysize=20.0cm
           \epsffile{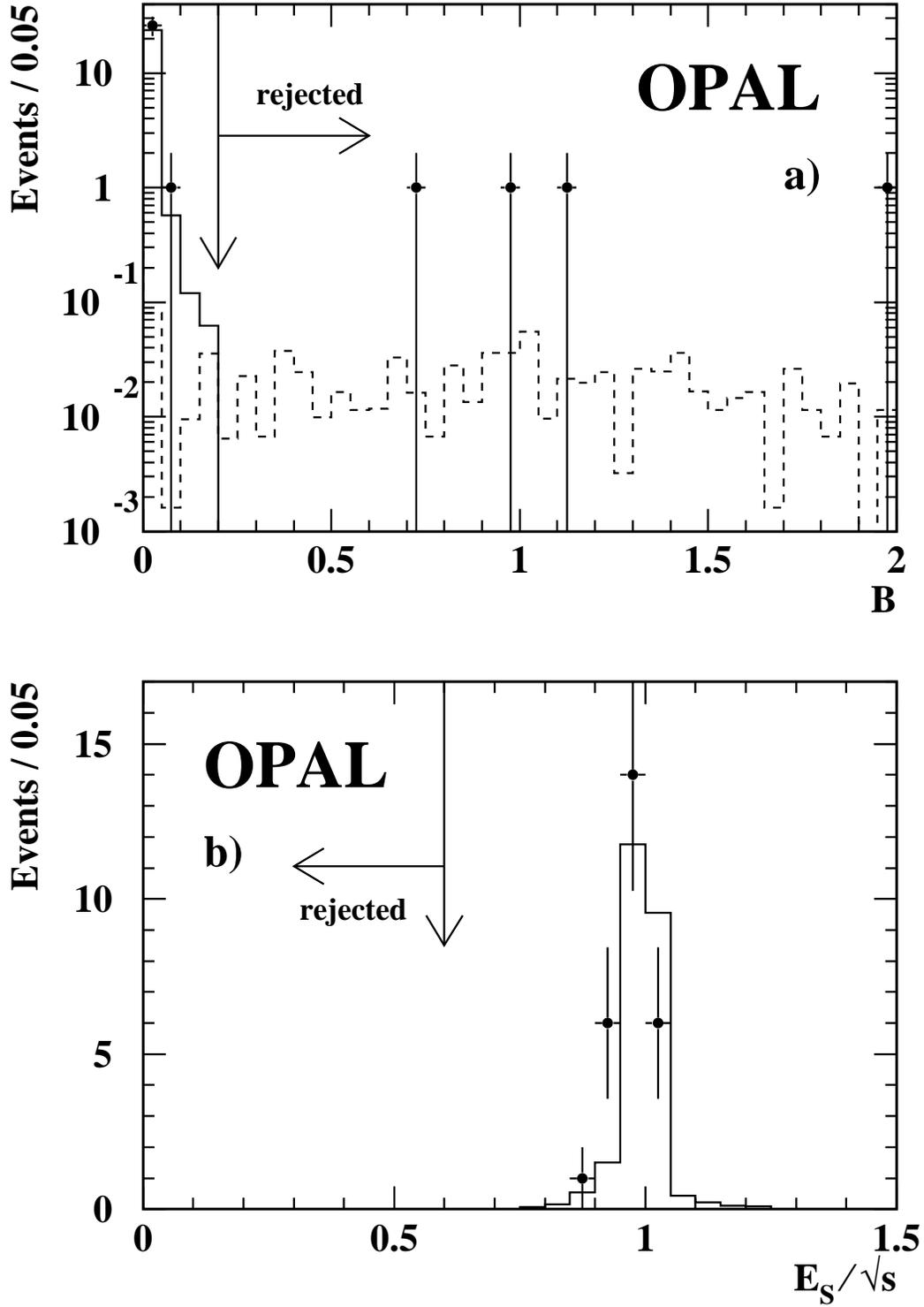}
           }\end{center}
\caption[ ]{
Event distributions for data in class \cb\ for $\sqrt{s} =$ 130, 136, 161 
and 172 GeV. 
Plot a) shows the distribution of the imbalance $\B$, a measure of the scaled
transverse momentum (for definition see Eq. \ref{dphi}) together with the 
selection cut.
Plot b) shows the scaled sum of both cluster energies plus E$_{\rm lost}$ 
after the cut on $\B$. The cut is indicated.
The points represent the data, the solid histogram the Monte Carlo expectation from 
$\eeggg$ and the dashed histogram the Monte Carlos expectation from
background (mainly $\nu\overline{\nu}\gamma\gamma$).}
\label{gg-g}
\end{figure}

Class \cb\ contains acollinear events ($\zeta > 10^{\circ}$) with exactly 
two observed photon candidates. Events of this class typically contain an
energetic photon that escapes detection near the beam-pipe ($|\ct |>0.97$). 
If the polar angle of 
this photon is approximated as $|\ct| =1$, its energy, $E_{\rm lost}$ 
(Eq. \ref{elost}), can 
be estimated from the angles of the observed photons $\theta_1$ and $\theta_2$.
The energy sum $E_S$ is then defined as the sum of the two observed cluster 
energies and the lost energy.
\ba
E_{\rm lost} &=&  \sqrt{s} \; \left( 1 + \frac{\sin{\theta_1} + \sin{\theta_2}}
                       {|\sin{(\theta_1 + \theta_2)}|}\right)^{-1}  
 \label{elost} \\
E_S & = & E_1 + E_2 + E_{\rm lost}  
\ea
The imbalance $\B$, defined as
\be
\B = (\sin{\theta_1}+\sin{\theta_2}) 
\left| \cos{\left(\frac{\phi_1-\phi_2}{2}\right)}\right| ,
\label{dphi}
\ee
provides an approximate measure of the scaled transverse momentum of the 
event without using the cluster energies.
Fig. \ref{gg-g} shows the distributions of $\B$ and $E_S$. It can be seen
that the background (mainly $\nu\overline{\nu}\gamma\gamma$) is uniformly
distributed in $\B$ whereas the signal is peaked at low values.
Events are selected if $\B < 0.2$ and $E_S > 0.6  \sqrt{s}$. 
Since the angular definition discussed in section \ref{angdef} uses the two 
highest energy photons, events are rejected if $E_{\rm lost}$ exceeds 
the energy of either observed photon.

\begin{figure}[htbp]
   \begin{center} \mbox{
          \epsfysize=20.0cm
           \epsffile{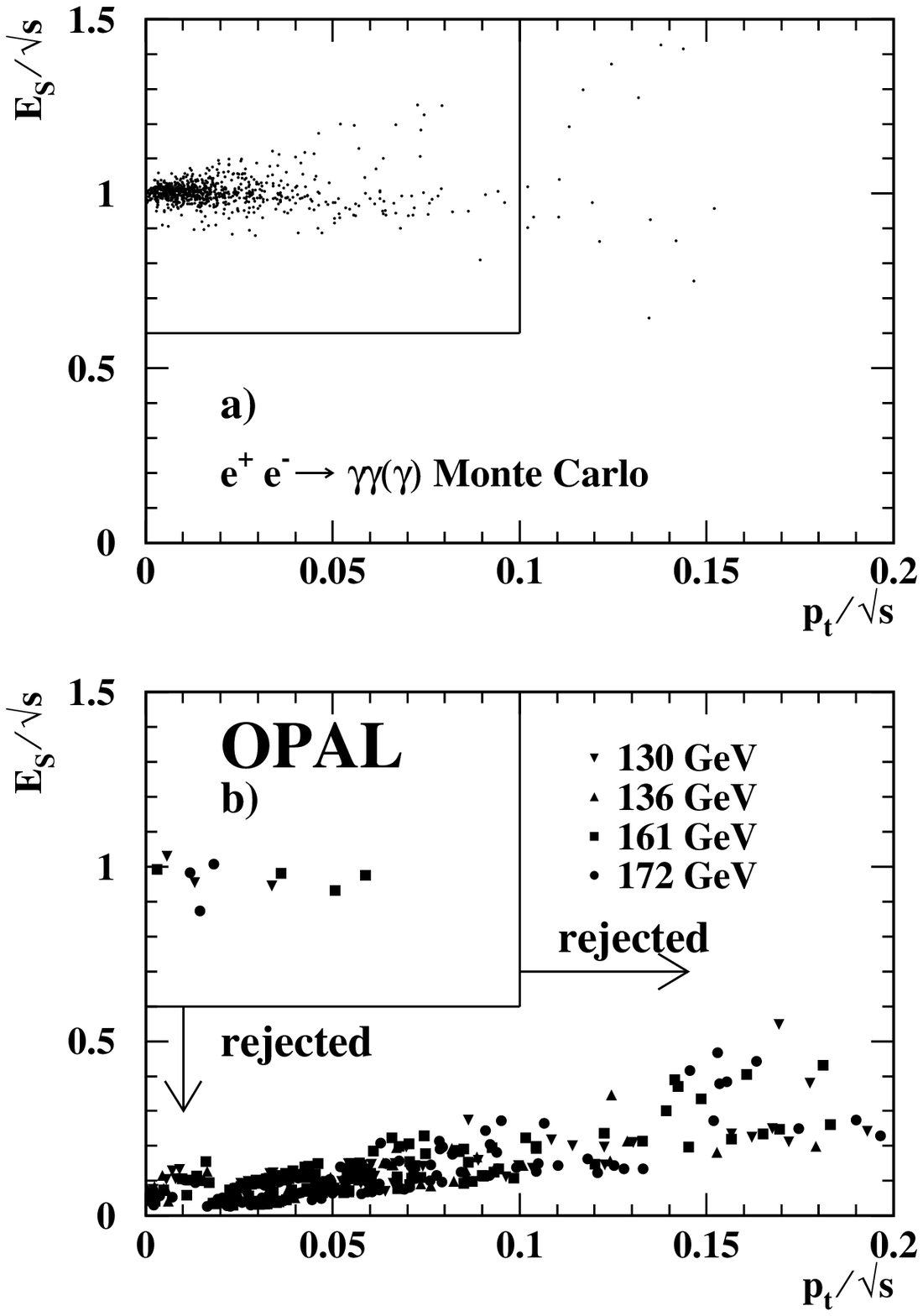}
           } \end{center}
\caption[ ]{The scaled energy sum versus the scaled 
transverse momentum for class \cc\ events, for a) $\eeggg$ Monte Carlo 
and b) the OPAL data. The box indicates the selected region. The 
background comes mainly from cosmic ray events.}
\label{pe}
\end{figure}

Class \cc\ contains acollinear events ($\zeta > 10^{\circ}$) having 
3 or more observed photon candidates. To calculate the transverse and 
longitudinal momenta ($\pt$, $\pl$) of the system the cluster energies
have to be used in addition to the photon angles. Since a non-zero longitudinal 
momentum could correspond to an additional photon along the beam direction, 
the energy sum $E_S$ is calculated as sum of the cluster energies $E_i$ 
and $\pl$: 
\be E_S =  \sum_{i=1}^n E_i + \pl . \ee
Fig. \ref{pe} shows the distribution of $E_S / \sqrt{s}$
versus $p_{\rm t}/\sqrt{s}$ for $\eeggg$
Monte Carlo and for the data. In the data the $\eeggg$ events are clearly 
separated from the background by the fact that they have small transverse 
momenta and an energy sum around the centre-of-mass energy. The main part of
the background originates from cosmic ray events without hits in the 
muon chambers. The selection requirements $E_S > 0.6 \sqrt{s}$ and  
$\pt < 0.1 \sqrt{s}$ easily reject these events.

No event with more than three clusters is observed.
The angle sum $\sum \alpha$ is used to identify planar three-photon events:
\be \sum \alpha = \alpha_{ij} + \alpha_{ik} + \alpha_{kj},
\ee
with $\alpha_{ij}$ the angle between photons $i$ and $j$. 
Seven planar events with $\sum \alpha > 350^{\circ}$ are accepted as 
three-photon events and included in the sample of $\eeggg$.
Two other events are consistent with three detected photons and an
additional photon along the beam direction.

The kinematic requirements used to select each of the event 
classes are summarised in Tab. \ref{cuts}. 

\begin{table}[h] \begin{center}
\begin{tabular}{|lr|r@{ }c@{ }l|} \hline
\multicolumn{2}{|l|}{Event class} & \multicolumn{3}{c|}{Requirements} \\ \hline
 all     &   & $\cte$&$<$&$ 0.9  $  \\\hline
\ca      &   & $\zeta$&      $<$ &$10^{\circ}$ \\
         &   & $E_S$&     $>$ &$0.6  \sqrt{s}$ \\\hline
\cb      &   & $\zeta$&      $>$ &$10^{\circ}$ \\
         &   & $E_S$&     $>$ &$0.6  \sqrt{s}$ \\
          &  & $E_1 , E_2$&  $>$ &$E_{\rm lost} $ \\
         &   & $\B$&          $<$ &$0.2$ \\
         &   & \multicolumn{3}{l|}{2 photon candidates}\\\hline
\cc      &   & $\zeta $&     $>$ &$10^{\circ}$ \\
         &   & $E_S $&    $>$ &$0.6  \sqrt{s}$ \\
         &   & $p_t $&       $<$ &$0.1  \sqrt{s}$ \\
         &   & \multicolumn{3}{l|}{$\geq$ 3 photon candidates}\\
         \cline{2-5}
          & planar & $ \sum \alpha$ & $>$ & $350^{\circ}$ \\
         \cline{2-5}
          & nonplanar & $ \sum \alpha$ & $<$ & $350^{\circ}$ \\
         \hline
\end{tabular}
\caption{Summary of the kinematic cuts. For definition of the variables see the text.}
\label{cuts}\end{center}
\end{table}

\section{Corrections and systematic errors}
\label{syserr}

Since the deviations from QED (Eqs. \ref{lambda} - \ref{estar}) are given with
respect to Born level,
the observed angular distributions  need to be corrected to Born level.
The effect of radiative corrections to the Born level calculation is quantified 
by $\R$, the ratio of the angular distribution of $\eeggg$ Monte Carlo and the 
Born cross-section as shown in Fig. \ref{mctest}:
\be \R = \xmc \!(\cte) \; \left/ \; \xb \right. .\ee
The ratio $\R$ is used to correct the data bin by bin to the Born level. A
1\%\ error on the total cross-section from higher order effects is assumed.
No error on the slope of the distribution is included in the results.
Since the $\O (\alpha^3)$ radiative corrections are small, the $\O (\alpha^4)$ 
effects are assumed to be negligible.

The efficiency and angular resolution of the reconstruction is determined 
using a Monte Carlo sample with full detector simulation. 
The efficiency is reasonably constant for $\cte < 0.9$, but drops rapidly for 
$\cte > 0.9$. The overall efficiency for $\cte < 0.9$ is 91.9\% with a maximum 
of 95\% in the barrel of the detector.
A polynomial parametrisation $\E (\cte)$ is used for the efficiency correction.
Due to uncertainties of the photon conversion probability and the 
Monte Carlo statistics a 1\%\ systematic error is assumed for the 
efficiency.
The agreement between generated and reconstructed angles is 
very good. An angular resolution of $0.3^{\circ}$ 
full width at half maximum is obtained.
Background is studied using Monte Carlo events from the processes shown in
Tab. \ref{tabbg}. The expected ratio of background to signal is less than
0.4\%\ and is neglected.

\begin{table}[h] \begin{center}
\begin{tabular}{|l|c|}\hline
Process Generator & Background events\\ \hline
$\epem\to\epem$ (BHWIDE)        & $<$ 0.24 \\
$\epem\to\epem$ (TEEGG)         & $<$ 0.67 \\
$\epem\to\bar{\nu}\nu\g\g $     & $<$ 0.02 \\
$\epem\to\mu^+\mu^- $           & $<$ 0.04 \\
$\epem\to\tau^+\tau^- $         & $<$ 0.05 \\
$\rm \epem\to\bar{q}q$          & $<$ 0.02 \\\hline
\end{tabular} \end{center}
\caption[ ]{Estimated 95 \%\ CL upper limits for expected background processes
from Monte Carlo at $\sqrt{s} =$ 130 - 172 GeV.}
\label{tabbg}
\end{table}

The probability that a signal event is rejected by the 
neutral event selection due to random instrumental 
background causing a veto is studied with randomly-triggered events. 
For the single veto the probability is $4 \times 10^{-4}$ and it is 
$1 \times 10^{-4}$ for both the double veto and the track veto for.
The small overall veto probability of  $5 \times 10^{-4}$ 
is therefore neglected.
The systematic errors on the total cross-section are summarized in Tab. 
\ref{syerr}. 

\begin{table}[h]
\begin{center}
\begin{tabular}{|l|c|} \hline
     & Uncertainty \\\hline
Luminosity    &  0.5\% \\
Radiative correction $\R$ & 1.0\% \\ 
Selection efficiency $\E$ & 1.0\% \\
Background  & $<$ 0.4\% \\ \hline
Total & 1.6\% \\\hline
\end{tabular}  \end{center}   
\caption{Summary of systematic errors on the cross-section}
\label{syerr}
\end{table}

\section{Results}

\begin{table}[b]

\begin{center}
\begin{tabular}{|l|r@{$\pm$}l|r@{$\pm$}l|r@{$\pm$}l|r@{$\pm$}l|}
\hline
Energy $\sqrt{s}$ [GeV] & \multicolumn{4}{c|}{130} & \multicolumn{4}{c|}{136}\\
\hline
 & \multicolumn{2}{c|}{Expected} & \multicolumn{2}{c|}{Observed}
 & \multicolumn{2}{c|}{Expected} & \multicolumn{2}{c|}{Observed}\\
\hline
Class \ca  & 34.3&0.9  &  \multicolumn{2}{c|}{33} 
&  30.1&0.8  & \multicolumn{2}{c|}{26}  \\
Class \cb  &  4.0&0.3  &   \multicolumn{2}{c|}{2}  
&   3.5&0.3  & \multicolumn{2}{c|}{3}  \\
Class \cc\ planar &  1.0&0.2  &   \multicolumn{2}{c|}{2}  
&   0.9&0.2  &  \multicolumn{2}{c|}{0}  \\ 
Class \cc\ nonplanar &  \multicolumn{2}{c|}{--}  &   \multicolumn{2}{c|}{0}  &  
 \multicolumn{2}{c|}{--}  &  \multicolumn{2}{c|}{0}  \\ \hline
$\rm\sigma_{tot}^{Born}$ & \multicolumn{2}{c|}{15.7} & 14.9&2.5
& \multicolumn{2}{c|}{14.3} & 12.2&2.2\\ \hline
\hline
Energy $\sqrt{s}$ [GeV] & \multicolumn{4}{c|}{161} & \multicolumn{4}{c|}{172}\\
\hline
 & \multicolumn{2}{c|}{Expected} & \multicolumn{2}{c|}{Observed}
 & \multicolumn{2}{c|}{Expected} & \multicolumn{2}{c|}{Observed}\\
\hline
Class \ca  & 85.8&0.6  &  \multicolumn{2}{c|}{90} 
 &  77.6&0.7  & \multicolumn{2}{c|}{75}  \\
Class \cb  &  9.0&0.2  &   \multicolumn{2}{c|}{8} 
 &   8.0&0.2  & \multicolumn{2}{c|}{14}  \\
Class \cc\ planar   &  1.9&0.1  &   \multicolumn{2}{c|}{3} 
 &   1.8&0.1  &  \multicolumn{2}{c|}{2}  \\ 
Class \cc\ nonplanar & \multicolumn{2}{c|}{--}  &   \multicolumn{2}{c|}{1}  &  
                       \multicolumn{2}{c|}{--}  &  \multicolumn{2}{c|}{1} \\ \hline
$\rm\sigma_{tot}^{Born}$ & \multicolumn{2}{c|}{10.2} & 10.9&1.1
 & \multicolumn{2}{c|}{9.0} & 9.7&1.0\\ \hline
\end{tabular}
\caption[ ]{Comparison of number of observed events and Monte Carlo prediction.
For nonplanar events no expectation is given, since the $\O (\alpha^3)$ Monte
Carlo does not include these events. The two observed class \cc\ nonplanar events
are kinematically 
compatible with a fourth photon along the beam-direction. In addition the total
cross section corrected to the Born level is given.}
\label{tabxsn}
\end{center}
\end{table}

\begin{figure}[htbp]
   \begin{center}
      \mbox{
          \epsfxsize=16.0cm
          \epsffile{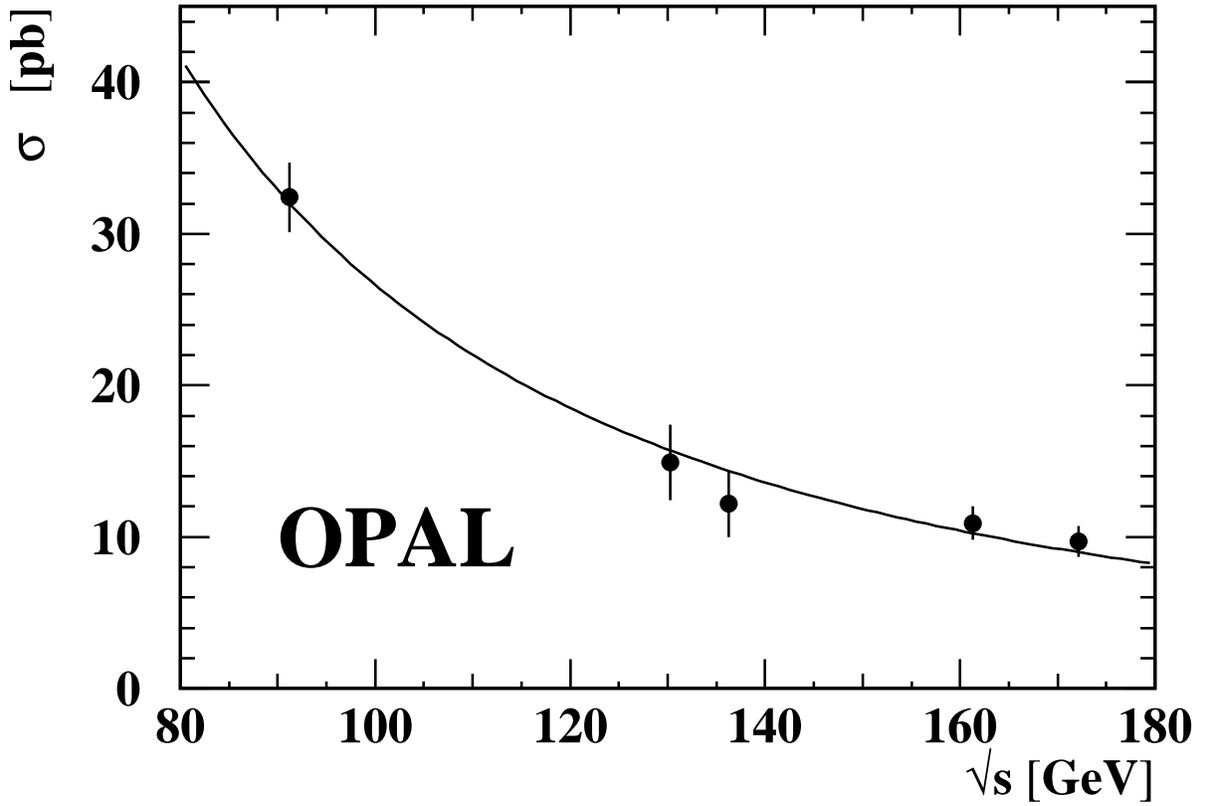}
           }
   \end{center}
\caption[ ]{Total cross-section for the process $\eegg$ with $\ct < 0.9$.
The data are corrected for efficiency loss and higher order effects and
correspond to a Born level measurement. The result at the Z$^0$ is taken from 
Ref. \cite{o1}.
The curve corresponds to the Born level QED prediction. }
\label{totx}
\end{figure}

In Tab. \ref{tabxsn} the numbers of observed events in each class are compared to the QED 
expectations. 
The derived total cross-section $\sigma$ in the range $\cte <$ 0.9 
is plotted in Fig. \ref{totx} as a function of the centre-of-mass energy. 
The numerical results for the cross-section are given in Tab. \ref{tabxsn}.
They are corrected for efficiency loss and $\O (\alpha^3)$
effects. All numbers agree well with QED expectations. 

\begin{figure}[htbp]
   \begin{center}
      \mbox{
          \epsfysize=20.0cm
          \epsffile{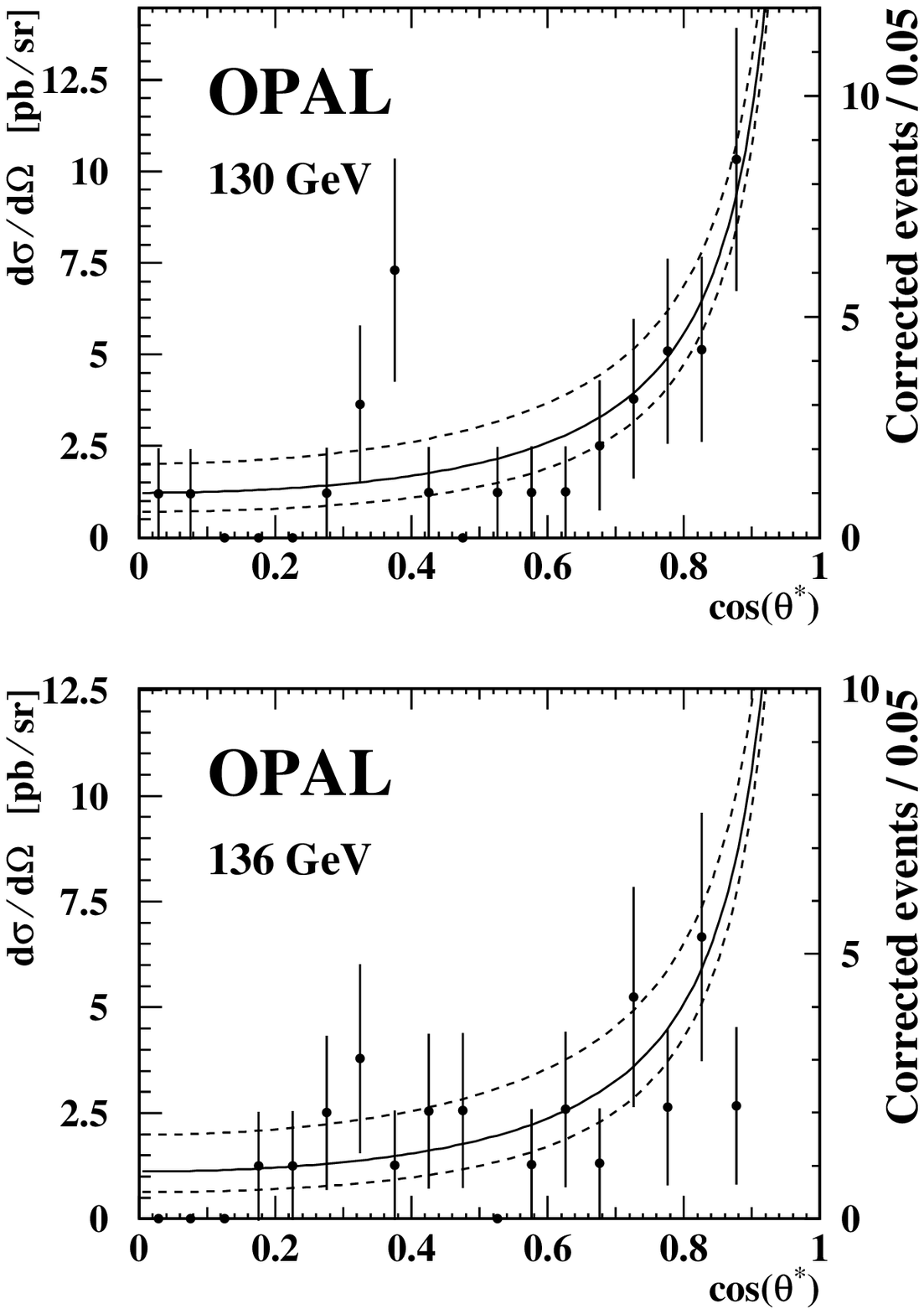}
           }
   \end{center}
\caption[ ]{The measured angular distribution for the process $\epem\to\g\g(\g)$
as selected in the three classes at $\sqrt{s} = $ 130 and 136 GeV.
The data points show the efficiency-corrected number of events; radiative 
corrections are also included.
The solid curve corresponds to the Born level QED prediction. 
The dotted lines represent 95\% CL intervals of the fit to the function $\xl$.  }
\label{wq1}
\end{figure}

\begin{figure}[htbp]
   \begin{center}
      \mbox{
          \epsfysize=20.0cm
          \epsffile{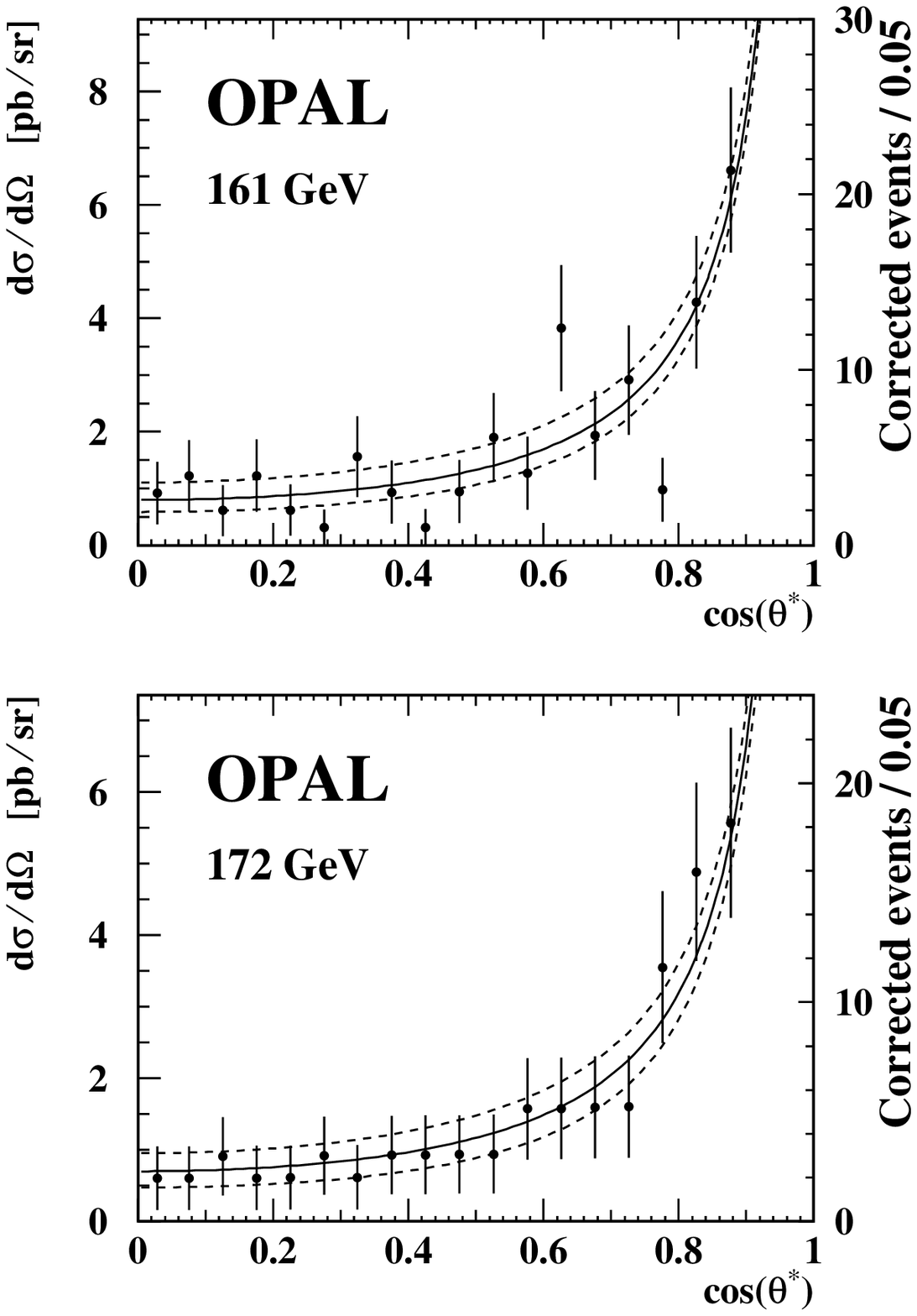}
           }
   \end{center}
\caption[ ]{The measured angular distribution for the process $\epem\to\g\g(\g)$
as selected in the three classes at $\sqrt{s} = $ 161 and 172 GeV.
The data points show the efficiency-corrected number of events; radiative 
corrections are also included.
The solid curve corresponds to the Born level QED prediction. 
The dotted lines represent 95\% CL intervals of the fit to the function $\xl$.  }
\label{wq2}
\end{figure}

The measured differential cross-sections at 130, 136, 161 and 172 GeV 
centre-of-mass energies are shown in Figs. \ref{wq1} and \ref{wq2} together 
with a fit of the function $\xl$ (Eq. \ref{lambda}).  
The fit to the distribution is performed using the binned log likelihood method. 
The likelihood function $\mathcal L$ is based on Poisson statistics 
and defined as:
\be {\mathcal L_i} = \frac{\mu_i^{n_i}}{n_i !} e^{-\mu_i} \ee
with $n_i$ the number of observed and $\mu_i$ the number of expected
events per $\ct$ bin $i$.
To determine $\mu_i$ the model dependent cross-section function $\xsn$ is not 
integrated over the bin. Instead a simple procedure is applied in which the 
central value $x_i$ of the bin is determined as defined in Ref. \cite{points}:
\be \xb(x_i) = \frac{1}{ x_u - x_l } \; \int_{x_l}^{x_u} \xb(y) dy  , \ee
where $x_l$ and $x_u$ are the lower and upper boundaries of the bin $i$.
In this way the differential function $\xsn$ can be directly compared to 
the integrated number of events presented as a histogram.
The mean efficiency $\E_i$ and radiative
corrections $\R_i$ are included in the expectation $\mu_i$ for each bin.
To allow the total number of expected events to vary within the systematic
error, a normalization factor $\epsilon$ is added: 
\be \mu_i = \epsilon \; \xsn(x_i) \; ( x_u - x_l )  \E_i \R_i L \ee
\be  \E_i = \frac{1}{n_i} \; \sum_{j=1}^{n_i} \E(\ct_j), \ee
where $L$ is the integrated luminosity and $\ct_j$ is the angle of the $j$-th 
event.
An estimator function $P$ is defined which includes a Gaussian term with mean 
1 and width  $\delta = 0.016$ (see Tab. \ref{syerr}) 
to account for the error of the normalization $\epsilon$. 
The routine MINOS \cite{minos}, which provides asymmetric
errors, is used to minimize $P$:
\ba P & = & \frac{(\epsilon - 1)^2}{\delta^2} + 
                 \sum_i - 2 \; \ln{{\mathcal L}_i} \nonumber\\
           & = & \frac{(\epsilon - 1)^2}{\delta^2} + 
                 \sum_i 2 \left( \mu_i - \; n_i \ln{\mu_i}\right) .
                     \ea

The fit is performed with two free parameters: 
the normalization $\epsilon$ and the model dependent parameter $\lambda$ 
(see Tab. \ref{result} and Eqs. \ref{lambda}, \ref{qed7} and
\ref{estar}). To obtain the limits at 95\% confidence level the 
probability is normalized to the physically allowed region, 
i.e. $\lambda_+ > 0$ and $\lambda_- < 0$ as described in Ref. \cite{pdg}. 

\begin{table}[b] 
\begin{center}
\renewcommand{\arraystretch}{1.4}
\setlength{\tabcolsep}{2mm}
\begin{tabular}{|ccc|r|r|} \hline
 & & & \multicolumn{2}{c|}{Fit result} \\
Model &  $\lambda$ & $\sqrt{s}$ [GeV] &
        \multicolumn{1}{c|}{$\lambda$} &
        \multicolumn{1}{c|}{$\epsilon$}   \\\hline
  &   & 130 &$ \left(4.3 {+20.6 \atop -18.0}\right)\cdot 10^{-10}$ GeV$^{-4}$& 
  $0.999 \pm 0.016$ \\
  &   & 136 &$ \left(7.1 {+19.3 \atop -16.7}\right)\cdot 10^{-10}$ GeV$^{-4}$& 
 $0.999 \pm 0.016$   \\
{\Large $\xl$} & 
{\large $ \pm 1 / \Lambda_{\pm}^4$ }             
      & 161 &$ \left(1.53{+5.12 \atop -4.70}\right)\cdot 10^{-10}$ GeV$^{-4}$ & 
$ 1.001 \pm 0.016$  \\[-0.5ex]
  &   & 172 &$ \left(-0.36{+4.13 \atop -3.76}\right)\cdot 10^{-10}$ GeV$^{-4}$& 
 $1.000 \pm 0.016$   \\
  &   & 130 - 172 &$ \left(0.74{+3.17 \atop -2.97}\right)\cdot 10^{-10}$ GeV$^{-4}$ & 
 $1.000 \pm 0.016$  \\
  \hline
{\large$\xq$} &  \rule[-2ex]{0ex}{6ex} 
{\large$|1 / \Lambda^6|$}  
       & 130 - 172 &$ \left(5.57{+35.9 \atop -47.0}\right)\cdot  10^{-18}$ GeV$^{-6}$ 
  &  $1.000 \pm 0.016$    \\
  \hline
{\large$\xe $} & \rule[-2ex]{0ex}{6ex}  
{\large$|1/M_{\rm e^*}^2|$} & 
   130 - 172 &$ \left(8.4{+11.0 \atop -27.9}\right)\cdot  10^{-6}$ GeV$^{-2} $  & 
  $1.000 \pm 0.016$  \\
  \hline  
\end{tabular}
\caption{Results for fit parameters $\lambda$ and $\epsilon$. For $\Lambda_{\pm}$ 
the results for all energies are shown seperately. The error on the normalisation
$\epsilon$ refects the assumed systematic error.}
\label{result}
\renewcommand{\arraystretch}{1.0}
\end{center}
\end{table}

\begin{figure}[htbp]
   \begin{center}
      \mbox{
          \epsfxsize=16.0cm
          \epsffile{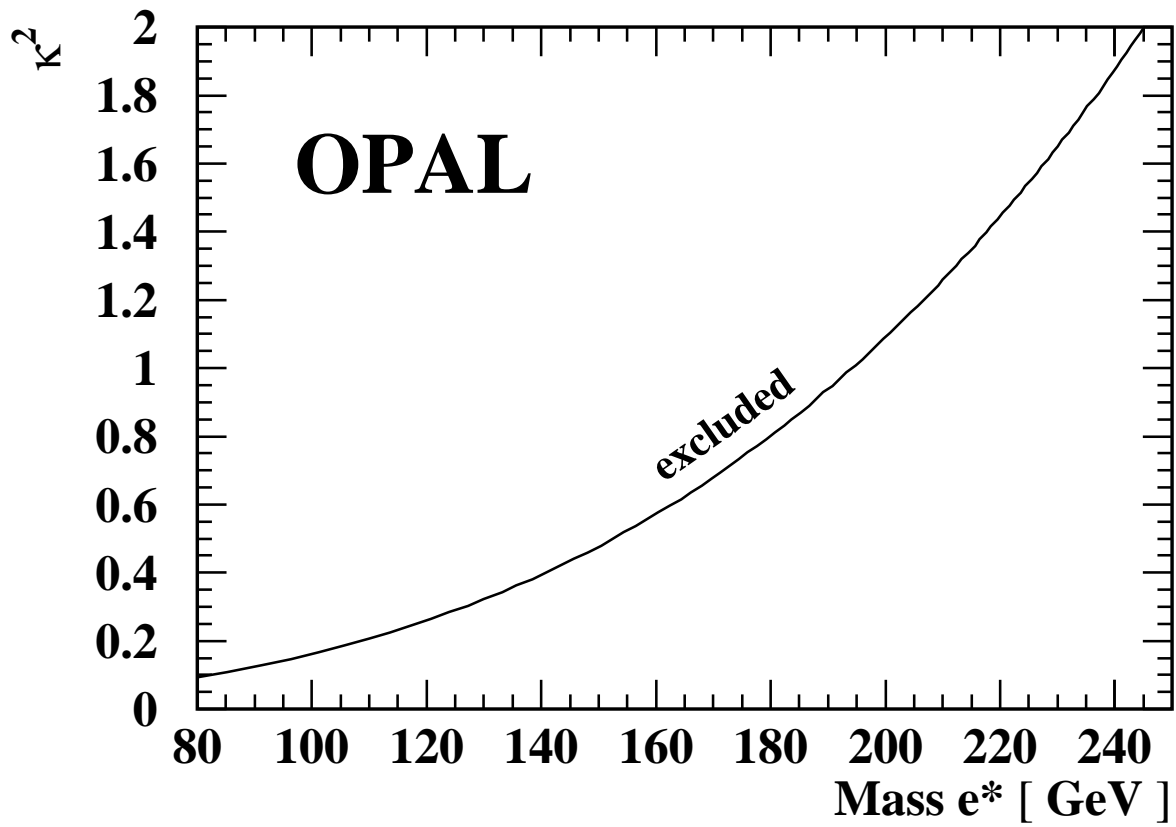}
           }
   \end{center}
\caption[ ]{Upper limit (95 \%\ CL) on the square of the coupling constant
$\kappa^2$ as a function of the mass of an excited electron $M_{\rm e^{\ast}}$.}
\label{elimit}
\end{figure}

Results for the different parameters are obtained from a simultaneous fit to the angular distibutions for each centre-of-mass energy. A fit is also 
performed for each centre-of-mass energy separately, with the results for 
$\Lambda_{\pm}$ given as an example in Tab. 
\ref{result}. The limits for the combined fit are summarised in Tab. 
\ref{tabsum}. To determine the limit on 
the mass of an excited electron $M_{\rm e^{\ast}}$ a fit is performed 
using $\xe$ (Eq. \ref{estar}). For the results given in Tab. \ref{result}
the coupling constant $\kappa$ for the ($\mathrm{e^*e\g}$)-vertex is fixed at 
$\kappa=1$. Fig. \ref{elimit} shows the upper limit (95 \%\ CL) on
$\kappa^2$ versus the mass of an excited electron $M_{\rm e^{\ast}}$.

The angular distributions for all energies agree well with the QED expectation. 
The lower limits 
obtained from the combined fit on $\Lambda_{\pm}$, $\Lambda$ and 
$M_{\rm e^{\ast}}$ are higher than 
existing published results using lower energies 
(see Ref. \cite{old}, \cite{o1}, \cite{a2} and \cite{l2})
and are in agreement with other results obtained at this 
centre-of-mass energy \cite{l3}. 
Previous limits on the excited electron mass with $\kappa=1$ are  
$M_{\rm e^{\ast}} > 129$ GeV \cite{l2}, 136 GeV \cite{a2} and 147 GeV \cite{graham}.

\begin{figure}[htbp]
   \begin{center} \mbox{
          \epsfxsize=16.0cm
           \epsffile{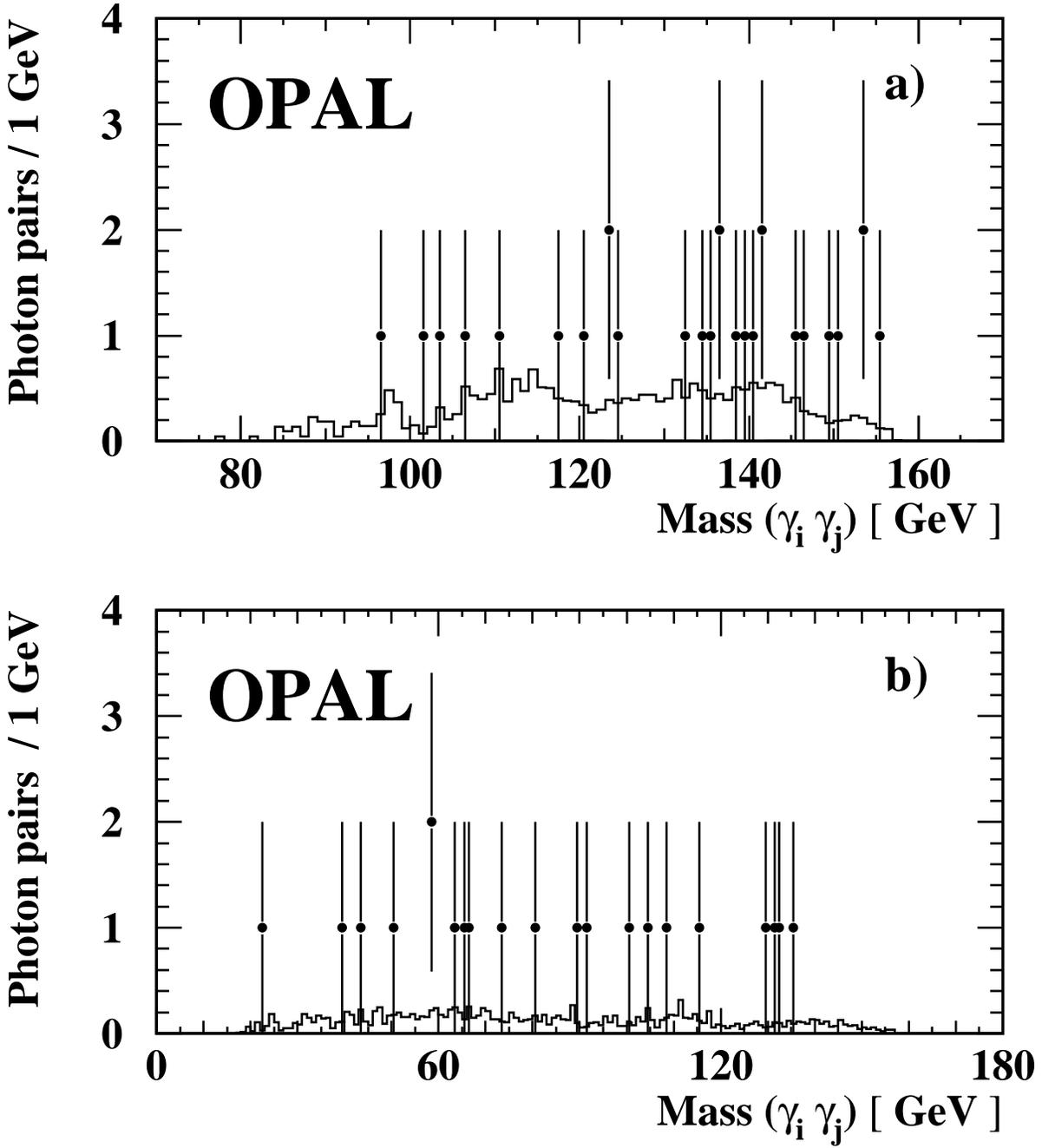}
           } \end{center}
\caption[ ]{The invariant mass of photon pairs for a)
class \cb\ events and b) class \cc\ events. The points are the data, 
the histogram the $\eeggg$ Monte Carlo expectation. 
There is one entry per event for class \cb\ events
and three entries per event for class \cc\ events.}
\label{pmass}
\end{figure}

\begin{figure}[htbp]
   \begin{center}
      \mbox{
          \epsfxsize=16.0cm
          \epsffile{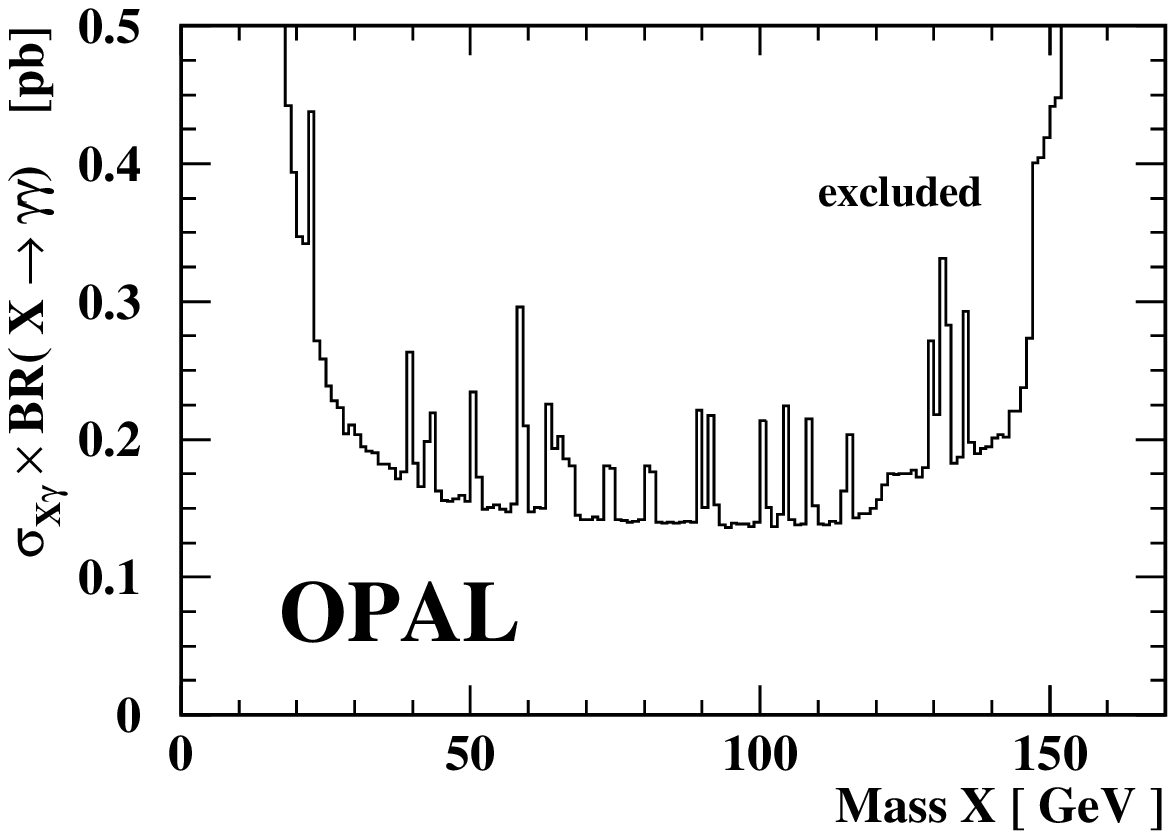}
           }
   \end{center}
\caption[ ]{Lower limits (95 \%\ CL) for the cross section times branching
ratio for the process $\rm \epem\to X \g$, $\rm X\to\g\g$ as a function
of the mass of the resonance X. The $\eeggg$ background is subtracted.
The step at 120 GeV comes from the phase space limit of the 130 and
136 GeV data.}
\label{mlimit}
\end{figure}

A resonance X produced in the process $\epem\to{\rm X}\g$ and decaying 
photonically ${\rm X} \to \g\g$ would be seen in the two photon invariant mass 
spectrum since this process leads to a three-photon final state without missing 
energy. This search has been performed previously at the Z$^0$-peak \cite{gres}
and at higher energies \cite{l2}.
The invariant mass of each photon pair is shown in Fig. \ref{pmass}
for all events of classes \cb\ and \cc .
There are three entries for events with three clusters.
Since the angular resolution is very precise,
the energies of the three photons are calculated from the angles assuming
three photon kinematics:
\be
E_k \propto \sin{\alpha_{ij}} \; ; \;  E_1 + E_2 + E_3 = \sqrt{s} ,
\ee
with $E_k$ the energy of one photon and $\alpha_{ij}$ the angle between the
two other photons. For class \cb\ events $|\ct |= 1$ is assumed for 
the unobserved photon. A typical mass resolution for photon pairs of
about 0.5 (0.7) GeV can be achieved for class \cc\ (\cb ).   
The distribution 
agrees well with the Monte Carlo expectation from the QED process $\eeggg$,
with no enhancement due to a resonance is observed. From the class \cc\ distribution
an upper limit on the total production cross-section times the photonic branching ratio of an isotropically 
produced resonance is calculated using the method of Bock \cite{bock}.
Combining the data of all centre-of-mass energies and subtracting the
$\eeggg$ background the limits shown in Fig. \ref{mlimit} are obtained.
The mass range is defined by the phase space of the selection and
limited due to the acollinearity restriction.

\begin{table}[h]
\begin{center}
\renewcommand{\arraystretch}{1.3}
\setlength{\tabcolsep}{4mm}
\begin{tabular}{|l|cccccc|}\hline
Parameter & $\Lambda_+$ & $\Lambda_-$ & 
  $\Lambda_6$ & $\Lambda_7$ & $\Lambda_8$ & $M_{\rm e^{\ast}}$ \\ 
  {[GeV]} &  195 & 210 & 793 & 483 & 15.5 &  194  \\
\hline
\end{tabular}
\renewcommand{\arraystretch}{1.0}
\caption[ ]{Summary of 95\% CL lower limits obtained from
the combined fit to the $\sqrt{s} =$ 130, 136, 161 and 172 GeV angular 
distributions. The results are for the cut-off parameters $\Lambda_{\pm}$ 
and mass scales $\Lambda$ according to {\small QED+6, QED+7} and
{\small QED+8} expectation (Eqs. \ref{qed6} - \ref{qed8}). $\Lambda_6$ and
$\Lambda_8$ are derived from $\Lambda_+$ and $\Lambda_7$ respectively.
The lower limit for the mass of an excited electron is also determined
with the coupling constant $\kappa$ assumed to be $\kappa = 1$. }
\label{tabsum}
\end{center}
\end{table}

\section{Conclusions}

The QED process $\eeggg$ has been studied using data taken with the OPAL
detector at LEP energies above the Z$^0$ resonance.
Both the angular distributions and the total cross-section measurement
agree well with QED predictions.
Limits are set on cut-off parameters, mass scales for 
contact interactions ($\g\g\epem$) and for non-standard $\g\epem$ couplings,
as well as on the mass of an excited electron coupling to $\mathrm{e\g}$.
These limits are listed in Tab. \ref{tabsum}.
In the $\gamma\gamma$ invariant mass spectrum 
of events with three final state photons, no 
evidence is found for a resonance X decaying to $\g\g$.
No photonic event with four or more detected photons is observed.

\section{Acknowledgements}

We particularly wish to thank the SL Division for the efficient operation
of the LEP accelerator at all energies
 and for
their continuing close cooperation with
our experimental group.  We thank our colleagues from CEA, DAPNIA/SPP,
CE-Saclay for their efforts over the years on the time-of-flight and trigger
systems which we continue to use.  In addition to the support staff at our own
institutions we are pleased to acknowledge the  \\
Department of Energy, USA, \\
National Science Foundation, USA, \\
Particle Physics and Astronomy Research Council, UK, \\
Natural Sciences and Engineering Research Council, Canada, \\
Israel Science Foundation, administered by the Israel
Academy of Science and Humanities, \\
Minerva Gesellschaft, \\
Benoziyo Center for High Energy Physics,\\
Japanese Ministry of Education, Science and Culture (the
Monbusho) and a grant under the Monbusho International
Science Research Program,\\
German Israeli Bi-national Science Foundation (GIF), \\
Bundesministerium f\"ur Bildung, Wissenschaft,
Forschung und Technologie, Germany, \\
National Research Council of Canada, \\
Hungarian Foundation for Scientific Research, OTKA T-016660, 
T023793 and OTKA F-023259.\\

\clearpage

%

%

\begin{thebibliography}{20}
\bibitem{graham}OPAL Collaboration, G. Alexander et al., Phys. Lett. {\bf B377}
                (1996) 222
\bibitem{misspt}OPAL Collaboration, K. Ackerstaff et al., Phys. Lett. {\bf B391}
                (1997) 210 
\bibitem{det}OPAL Collaboration, K. Ahmet et al., Nucl. Instr. and Meth.
             {\bf A305} (1991) 275
\bibitem{mcgen}F.A. Berends and R. Kleiss, Nucl. Phys. {\bf B186} (1981) 22
\bibitem{mcbh}S. Jadach et al., Phys. Lett. {\bf 390} (1997) 298
\bibitem{mcte}D. Karlen, Nucl. Phys. {\bf B289} (1987) 23
\bibitem{mcnn}G. Montagna et al., Nucl. Phys. {\bf B452} (1996) 161
\bibitem{mctt}S. Jadach et al., Comp. Phys. Comm. {\bf 66} (1991) 276
\bibitem{mcmh}T. Sj\"ostrand and M. Bengtsson, Comp. Phys. Comm. {\bf 43} (1987) 367 \\
              T. Sj\"ostrand, Comp. Phys. Comm. {\bf 39} (1986) 347
\bibitem{mcdet}OPAL Collaboration, J. Allison et al., Nucl. Instr. and Meth.
             {\bf A317} (1992) 47
\bibitem{bg}I. Harris and L.M. Brown, Phys. Rev. {\bf 105} (1957) 1656 \\
            F.A. Berends and R. Gastmans, Nucl. Phys. {\bf B61} (1973) 414
\bibitem{drell}S.D. Drell, Ann. Phys. {\bf 4} (1958) 75
\bibitem{eboli}O.J.P. \'{E}boli, A.A. Natale and S.F. Novaes,
            Phys. Lett. {\bf B271} (1991) 274
\bibitem{litke}A. Litke, Ph.D.Thesis, Harvard University, unpublished (1970)
\bibitem{points}G.D. Lafferty and T.R. Wyatt, Nucl. Instrum. Meth. {\bf A355} (1995) 541
\bibitem{minos}MINUIT Reference Manual, F. James and M. Roos, CERN Program Library {\bf D506}
\bibitem{pdg}Review of Particle Physics, R.M. Barnett et al., Phys. Rev. {\bf D54} (1996) 1
\bibitem{old}PLUTO Collaboration, C. Berger et al., Phys. Lett. {\bf B59} (1980) 87
\\ JADE Collaboration, W. Bartel et al., Z. Phys. {\bf C19} (1983) 197
\\ MARKJ Collaboration, B. Adeva et al., Phys. Rev. Lett. {\bf 53} (1984) 134
\\ TASSO Collaboration, M. Althoff et al., Z. Phys. {\bf C26} (1984) 337 
\\ CELLO Collaboration, H.J. Behrend et al., Phys. Lett {\bf B168} (1986) 420
\\ HRS Collaboration, M. Derrick et al., Phys. Rev. {\bf D34} (1986) 3286
\\ MAC Collaboration, E. Fernandez et al., Phys Rev. {\bf D35} (1987) 1
\\ AMY Collaboration, H.J. Kim, et al., KEK preprint {\bf 89-52} (1989)
\\ VENUS Collaboration, K. Abe et al., Z. Phys. {\bf C45} (1989) 175
\\ TOPAZ Collaboration, K. Shimozawa et al., Phys. Lett. {\bf B284} (1992) 144
\\ ALEPH Collaboration, D. Buskulic et al., Z. Phys. {\bf C59} (1993) 215
\\ DELPHI Collaboration, P. Abreu et al., Phys. Lett. {\bf B268} (1991) 296
\\ L3 Collaboration, O. Adriani et al., Phys. Lett. {\bf B288} (1992) 404
\bibitem{o1} OPAL Collaboration, M.Z. Akrawy et al., Phys. Lett {\bf B257} (1991) 531
\bibitem{a2} ALEPH Collaboration, D. Buskulic et al., Phys. Lett. {\bf B384} (1996) 333
\bibitem{l2} L3 Collaboration, M. Acciarri et al., Phys. Lett. {\bf B384} (1996) 323
\bibitem{l3}L3 Collaboration, M. Acciarri et al., CERN-PPE/97-77,
            submitted to Phys. Lett.
\bibitem{gres} OPAL Collaboration, P.D. Acton et al., Phys. Lett {\bf B311} (1993) 391 \\
               L3 Collaboration, M. Acciarri et al., Phys. Lett. {\bf B345} (1995) 609
\bibitem{bock}P. Bock, 
{\it Determination of Exclusion Limits for Particle Production Using Different 
     Decay Channels with Different Energies, Mass Resolutions and Backgrounds},
     submitted to Nucl. Instrum. Meth. (1997) 
\end{thebibliography}
\end{document}